\documentstyle[natbib,color,amssymb,endnotes,doublespace,verbatim,times,12pt]{article}
\let\footnote=\endnote

\begin{document}
\title{The Personalized A-Theory of Time and Perspective\footnote{This
    paper will appear in {\em dialectica}.}}
\author{Vincent Conitzer\thanks{Departments of Computer Science, Economics,
    and Philosophy, Duke University, Box 90129, Durham, NC 27708, USA;
    Email: vincent.conitzer@duke.edu}}
\date{}
\maketitle
\begin{abstract}
  A-theorists and B-theorists debate whether the ``Now'' is metaphysically
  distinguished from other time slices.  Analogously, one may ask whether
  the ``I'' is metaphysically distinguished from other perspectives.  Few
  philosophers would answer the second question in the affirmative.  An
  exception is Caspar Hare, who has devoted two papers and a book to
  arguing for such a positive answer.  In this paper, I argue that those
  who answer the first question in the affirmative -- A-theorists -- should
  also answer the second question in the affirmative.  This is because key
  arguments in favor of the A-theory are more effective as arguments in
  favor of the resulting combined position, and key arguments
  against the A-theory are ineffective against the combined position.
\end{abstract}

\section{Introduction}
In a series of unconventional but lucid works, Caspar Hare has laid out and
defended a theory of {\em egocentric presentism} (or, in his more recent
work, {\em perspectival realism}), in which a distinguished individual's
experiences are {\em present} in a way that the experiences of others are
not~\citep{Hare07:Self,Hare09:On,Hare10:Realism}.  Closely related ideas
appear in the writings of others.  One example is~\citet{Valberg07:Dream}'s
notion of the ``personal horizon,'' especially considering his discussion
of ``the truth in solipsism'' and his insistence that ``my'' horizon is
really ``the'' (preeminent) horizon.  \citet{Merlo16:Subjectivism}'s
``subjectivist view of the mental'' is arguably even more closely related;
he argues that ``one's own mental states are metaphysically privileged
vis-\`a-vis the mental states of others'' and discusses in detail the
relationship of his view to Hare's.  As another example, in a review of
``The Character of Consciousness''~\citep{Chalmers10:Character},
\citet{Hellie13:Against} argues that this work fails to do justice to the
{\em embedded point of view} aspect of consciousness.  He illustrates this
with what he calls a ``vertiginous question'': why, of all subjects, is
{\em this} subject (the one corresponding to the human being Benj Hellie)
the one whose experiences are ``live''?  In other
work~\citep{Conitzer19:Puzzle}, I explore whether the ``liveness'' of one
particular perspective is a {\em further fact} -- a fact that does not
follow logically from the physical facts of the world -- by considering the
analogy to looking in on a simulated world through a virtual reality
headset: besides the computer code that determines the physics of the
simulated world, there must be additional code that determines which
simulated agent's perspective to show on the headset.

In any case, Hare's exposition of these ideas is clearest for the present
purpose, so I will focus on it.  In an effort, possibly with limited
success, to avoid misrepresenting his position, as well as to clarify the
relation to other work, let me introduce my own terminology.  Let us refer
to the theory that states that there is a metaphysically (rather than
merely epistemically) distinguished\footnote{Throughout the paper, I will
  be deliberately noncommittal about the exact nature of such metaphysical
  distinction.  The reason is that the arguments presented here do not
  depend on what this distinction consists in. In the analogous case of a
  metaphysically distinguished {\em time} (rather than a metaphysically
  distinguished subject), by not committing to any particular
  interpretation, I can simultaneously address all varieties of A-theorists
  -- presentists, moving-spotlight theorists, growing-block theorists,
  etc. -- even though they disagree about the exact nature of the Now's
  metaphysical distinction.  Of course, there is disagreement even about
  how to define the individual varieties. \citet{Deasy15:What} discusses
  this at length, and proposes to define each of the main varieties as the
  conjunction of the A-theory (which he takes to mean ``There is an
  absolute, objective present instant'') and a proposition about whether
  things begin and/or cease to exist.  For example, for the growing-block
  theorist, that proposition is ``Sometimes, something begins to exist and
  nothing ever ceases to exist.''  While the distinctions between the
  various definitions are significant, again, my aim is to steer clear of
  this debate here and stick to arguments that work for any of these
  definitions. The same is true for the case of a metaphysically
  distinguished subject.}  {\em I} (or {\em Self}\footnote{Again, what
  exactly the distinguished entity is -- a human being, a brain, an
  experience -- is not essential to my arguments, so I will remain
  deliberately noncommittal.}) as the $\alpha$-theory. The intent is to
emphasize the analogy with how the A-theory~\citep{McTaggart1908:Unreality}
states that there is a metaphysically distinguished {\em
  Now}.\footnote{\label{fo:facts}Is a commitment to a distinguished {\em
    Now} what defines the A-theory, or is it a commitment to tensed facts?
  (And in the latter case, should the $\alpha$-theory's defining commitment
  instead be to first-personal facts?)  To the extent that these
  commitments are not equivalent, in this paper, I will stick with the
  commitment to a distinguished {\em Now} (or {\em I}), as others have done
  -- e.g., \citet[page 89]{Cameron15:Moving}.  For what it is worth, while
  a detailed analysis is outside the scope of this paper, I believe that
  they are in fact equivalent.  I believe that a distinguished {\em Now}
  implies tensed facts, such as the fact that today is July 3, 2019.  The
  other direction is perhaps more controversial, but I believe it holds as
  well: tensed facts such as the fact that today is July 3, 2019
  distinguish a specific time, to which we may refer as the {\em Now}.  A
  theory such as fragmentalism~\citep{Fine05:Tense} might be used to
  dispute the second direction: if we consider {\em all} tensed facts,
  including those for other times, then no specific time is distinguished.
  But, of course, the set of all tensed facts taken together is full of
  contradictions, as it also contains, for example, the fact that today is
  not July 3, 2019.  Avoiding such contradictions means restricting
  attention to a consistent fragment -- but this in turn distinguishes a
  specific time.  For further discussion of problems that fragmentalism
  faces, see \citet[pages 86-102]{Cameron15:Moving}.}  Similarly, I will
refer to the theory that contradicts the existence of any metaphysically
distinguished {\em I} as the $\beta$-theory.  Hare is thus defending the
$\alpha$-theory.  It is not entirely clear to me whether the specific
version he defends is intended to be analogous to presentism (or actualism
-- I will refrain from discussing modality in this paper, but the parellels
between time/subjectivity and modality are well
recognized~\citep{Prior77:Worlds,Bergmann99:Serious}), or rather to
something like a spotlight theory (or possibilism).  In fact, his writing
suggests different answers to this question in different places, and I will
not attempt to resolve this small mystery here.

Others have commented on the idea of a metaphysically distinguished {\em I}
-- or, similarly but not equivalently,\footnote{For a discussion of the
  differences and their implications, in the related context of the
  Lewisian and Quinean accounts of centered worlds,
  see~\citet{Liao12:What}.} a metaphysically distinguished {\em Here} -- in
the context of the philosophy of time.  (While the differences between a
metaphysically distinguished {\em I} and a metaphysically distinguished
{\em Here} will not matter for some of the arguments presented in this
paper, it is useful to note that, in the context where a distinguished {\em
  I} is combined with a distinguished {\em Now}, the combination of these
two immediately implies a distinguished {\em Here} as well -- namely, the
location of the distinguished individual at the distinguished
time.\footnote{The combination similarly implies a distinguished
  observational frame of reference corresponding to the distinguished
  individual's state of motion. All of this does, of course, require the
  distinguished individual to be spatially located and to move through time
  and space, rather than, say, an immaterial soul or something existing for
  only an instant.})  However, they have usually dismissed it rather
quickly, in order to move on with the case of a metaphysically
distinguished {\em Now} (whether or not they support the latter).  For
example,~\citet[page 422]{Zimmerman05:ATheory} writes:
\begin{quote}
  An egocentric analogue of actualism (`personalism', to steal and abuse a
  term) is very hard to imagine. Perhaps there is some kind of
  not-merely-epistemological solipsism that would qualify. In any case,
  only the maniacally egocentric could be this sort of personalist.
\end{quote}
Further back,~\citet[page 458]{Williams51:Myth} writes:
\begin{quote}
  Perhaps there exists an intellectualistic solipsist who grants the
  propriety of conceiving a temporal stretch of events, to wit, his own
  whole inner biography, while denying that the spatial scheme is a literal
  truth about anything. Most of the disparagers of the manifold, however,
  are of opposite bias. Often ready enough to take literally the spatial
  extension of the world, they dispute the codicil which rounds it out in
  the dimension of time.
\end{quote}
\citet{Fine05:Tense} treats the case of first-personal realism in detail,
but advocates for adopting a nonstandard variety of realism, either taking
reality to be relative to a standpoint, or (his preferred option)
considering it to be fragmented.\footnote{\citet{Lipman15:On} discusses
  fragmentalism in more detail.}  He notes:
\begin{quote}
  It has seemed evident that, of all the possible worlds, the actual world
  is privileged; it is the standpoint of reality, as it were, and the facts
  that constitute reality are those that obtain in this world. On the other
  hand, if we ask, in the first-personal case whether we should be a
  nonstandard realist (given that we are going to be first-personal
  realists in the first place), then the answer to most philosophers has
  seemed to be a clear `yes'.  It has seemed metaphysically preposterous
  that, of all the people there are, I am somehow privileged - that my
  standpoint is {\em the} standpoint of reality and that no one else can
  properly be regarded as a source of first-personal facts.  The case of
  time is perplexing in a way that these other cases are not.
\end{quote}

I believe that there is value in exploring the $\alpha$-theory more
thoroughly, rather than dismissing it summarily for being repugnant in one
way or another.  The words ``egocentrism'' and ``solipsism'' are both
loaded with too much baggage.  While ``egocentric,'' taken literally, aptly
describes the $\alpha$-theory, the common interpretation of the word
carries various negative connotations, and it is not clear to me that these
are fair to apply to every possible $\alpha$-theorist. Just as A-theorists
can take great interest in times other than their own (otherwise why would
they bother to write papers?), the $\alpha$-theorist can presumably take
great interest in people other than herself.\footnote{It should be noted
  here that, on the face of it,~\citet{Hare07:Self,Hare09:On} does
  introduce his theory to justify placing greater weight on oneself than on
  others in making decisions.  However, he also points out that the
  (distinguished) presence of an experience is only one factor in making
  decisions (``It is better that there be present suffering from a hangnail
  than absent suffering of leg-crushing.'').  Perhaps more importantly, key
  examples that Hare uses in these works to support his theory are
  preferential in nature, such as an example where one knows that CJH
  (Hare) and Joe Bloggs have been in a train crash, CJH is about to have a
  painful operation, the subject knows he is one of these two but cannot
  remember which one, and so the subject hopes to not be
  CJH~\citep{Hare07:Self}.  Such preferential examples are quite helpful to
  illustrate and motivate these types of theories -- similar ones can be
  given to motivate the A-theory, as Hare does and others have done before
  him -- even if one does not wish to normatively endorse the preferences
  used in the example.  I will also discuss such examples later in this
  paper.}  The relation to solipsism is also not obvious.  Hare intends for
his theory to be only a weak and subtle version of solipsism that does not
deny the existence of others' consciousness~\citep[pages 41-46]{Hare09:On},
and others have granted him as much (e.g.,~\citet{Smith11:Inconsistency},
and Mark Johnston in the introduction
to~\citet{Hare09:On}).\footnote{Others have tried to distinguish between
  more and less defensible versions of solipsism along similar lines; a
  particularly notable example is~\citet{Valberg07:Dream}. Similar ideas
  also appear in~\citet{Johnston11:Surviving}.}

Indeed, a key point is that, just as there are multiple versions of the
A-theory, there are also multiple versions of the $\alpha$-theory, and
these vary in the status they accord to other individuals.  Perhaps more
importantly -- and this is the main focus of this paper -- something is
lost when attempting to study the A vs.~B question separately from the
$\alpha$ vs.~$\beta$ question; the two are very much interrelated.  To
illustrate this, consider a theory that allows a distinguished {\em I} that
is not alive at the time of the distinguished {\em Now}, thereby treating
the two types of distinction as independent.  Many of the arguments that I
give in what follows would do little to support such a theory.  Hence, in
what follows I will not take the $\alpha$A-theory -- the label that I will
use for a view that combines the $\alpha$-theory with the A-theory -- to
allow this possibility; what I have in mind is that a single {\em
  (living-)person-stage} is distinguished.  This interrelation is relevant
to the previous point.  For example, the $\alpha$A-theorist may accord to
other persons the same metaphysical status as she does to herself in past
and future time slices.

After presenting, for the sake of illustration, some versions of the
$\alpha$A-theory (Section~\ref{se:versions}), I will argue that key
arguments that have been given to support the A-theory support the
$\alpha$-theory just as well, and in fact support the combined
$\alpha$A-theory especially strongly (Section~\ref{se:favor}), placing the
onus on the $\beta$A-theorist to explain why she accepts the A-theory but
not the $\alpha$-theory.  (It would seem that most A-theorists, at least
publicly, are $\beta$A-theorists in my terminology.)  Specifically,
in~\ref{su:simpliciter} I will discuss the argument from presence {\em
  simpliciter}, and in~\ref{su:appropriate} the argument from the
appropriateness of sentiments such as those expressed by ``Thank goodness
that's over!''  I will also argue that some serious challenges that the
$\beta$A-theorist faces are much less problematic for the
$\alpha$A-theorist (Section~\ref{se:against}).  Specifically,
in~\ref{su:relativity} I will discuss the argument from special relativity,
in~\ref{su:direction} the argument that the direction of time may be a
local matter, in~\ref{su:rate} the argument that asks for the rate at which
time passes, and in~\ref{su:travel} the argument from time travel and
G\"odelian universes.

Overall, my main objective is to argue that the $\alpha$A-theory is
superior to the $\beta$A-theory.\footnote{Of course, to accept this
  conclusion, it is not necessary to agree with every single argument
  presented here.}  I would similarly argue that the $\alpha$A-theory is
superior to the $\alpha$B-theory, but I do not expect many to defend the
latter view.\footnote{In fact,~\citet[page 48]{Hare09:On} writes that ``If
  you think that theories that dignify a slice of history do not survive
  sustained critical inspection, then you can still be a
  four-dimensionalist egocentric presentist.  Indeed, I find that an
  attractive position.''  This may appear to put him in the $\alpha$B-camp.
  However, on the whole in this section on the relationship to positions in
  the philosophy of time~\citep[pages 46-50]{Hare09:On}, he is clear that
  egocentric presentism does not commit one to a particular view on time,
  while also stating that the moving-spotlight theory is the most analogous
  one.  Elsewhere~\citep{Hare10:Realism}, he writes ``If you find yourself
  sympathetic to [the central tense realist idea] then I recommend that you
  consider {\bf going the whole hog}, and becoming a perspectival realist''
  (emphasis in bold mine), which might be interpreted to imply that
  perspectival realism is a stronger position than tense realism.  In any
  case, as I hope will become clear from this paper, the $\alpha$A-theory
  does not at all require a dignified {\em slice} of history.}  This would
leave the $\alpha$A-theory and the $\beta$B-theory as the remaining
candidates.  The reader might expect that my next step will be simply to
argue that the $\alpha$-theory is so unappealing that we should accept the
$\beta$B-theory, and hence, {\em a fortiori}, the B-theory.  However, I
believe that that conclusion is too hasty; an effective discussion of the
relative merits of the $\alpha$A-theory and the $\beta$B-theory requires
arguments of a different type than what I will present here.  So, I will be
content to let both theories stand for now.

\section{Some versions of the $\alpha$-theory}
\label{se:versions}

The A-theory counts among its supporters presentists, moving-spotlight
theorists, and growing-block theorists.  Can we conceive of similar
distinctions among $\alpha$-theorists?  Rather than studying this in
isolation from the A vs.~B question, it seems more enlightening to ask what
natural versions of the $\alpha$A-theory there are.  (Common versions of
the A-theory and the B-theory can straightforwardly be reinterpreted as
versions of the $\beta$A-theory and the $\beta$B-theory.)  I will present
some versions in this section.  My aim here is not to defend specific
versions or to reach any definitive conclusion about which version is best.
I also make no claim that this list is exhaustive, though I believe that it
includes the versions that are most natural to discuss in the context of
the existing literature on the A-theory.  The aim of this exercise is
merely to clarify some aspects of the $\alpha$A-theory and prevent overly
narrow interpretations of it. Moreover, it will be helpful to refer to some
of these versions in what follows.  I will also contrast these versions
with some scenarios from the literature.

\begin{itemize}
\item {\bf Personalized presentism.} This is the most natural way to adapt
  presentism into an $\alpha$A-theory.  In this version, there is a single
  distinguished individual whose experience at a single distinguished point
  in time is, in some sense, ``present.'' (I hope that the intended meaning
  of ``presence'' is at least somewhat clear at this point; I will discuss
  it in more detail in~\ref{su:simpliciter}.)  Beyond this present
  experience, nothing exists. Or, perhaps, some part of the outer world can
  be granted some type of existence; but other experiences do not exist.
  However, presumably, the present experience can change (more on this
  below), just as presentists typically consider it possible for the Now to
  change.

\item {\bf Personalized moving spotlight.}  As in the classical
  moving-spotlight theory, a spotlight moves over the four-dimensional
  block universe, except now this spotlight shines on a single individual
  (or that individual's experience) at a single point in time.  For the
  personalized moving spotlight, it is less obvious how it moves (more on
  this below).

\item {\bf Personalized growing block.} In the classical growing block
  theory, time slices are added to the block that contain all the events in
  the universe at that point in time.  In the personalized growing block
  theory, only those parts of spacetime are added that are experienced by a
  distinguished individual (and, perhaps, their past light cones).
\end{itemize}
Every one of these versions of the $\alpha$A-theory leaves several
possibilities for {\em how} the point of present experiences -- the
``I-Now'' -- could change or move (if it changes or moves at
all).\footnote{The word ``I-Now'' sounds more mystical than I would like,
  but we will need such a word.  The word ``spotlight,'' when interpreted
  as shining on a single individual's experience at a single point in time,
  would give the right idea, except it seems to commit the discussion to a
  view that all of the four-dimensional spacetime block exists, but not all
  of it is illuminated.  While I do not want to dismiss such a view, in
  what follows we will not require this as an assumption. In contrast, the
  awkward word ``I-Now'' does not seem to rule out any of the
  possibilities. (Similarly, \citet{Hellie13:Against} uses ``me-now.'')}
These include the following variants:
 
\begin{itemize}
\item {\bf Single individual overall.} The I-Now moves along with a single
  individual throughout his or her lifetime. It is never associated with
  any other individual.

\item {\bf Changing individual ($\alpha$A-reincarnation).} At the end of
  the distinguished individual's lifetime, the I-Now jumps to another
  individual.  We can consider various subvariants. For example: (1) the
  I-Now cannot jump backwards in time; (2, a relativistic subvariant) the
  I-Now can jump anywhere that is outside of all the past light cones of
  points in spacetime that the I-Now occupied earlier; (3) the I-Now can
  jump anywhere it has not previously been; (4) the I-Now is not
  constrained in where it can jump.\footnote{The last two subvariants seem
    more difficult to reconcile with the personalized growing block theory,
    and might also have negative implications for free will.}

\item {\bf Rapidly changing individual.}  The I-Now can jump from one
  individual to another even before the former's demise, and then jump back
  to the previous individual as well.  We can consider the same subvariants
  as for $\alpha$A-reincarnation.
\end{itemize}

It is admittedly odd to propose all these different versions of the
$\alpha$-theory without making any serious attempt to justify them
individually or to claim to be exhaustive.\footnote{For example, perhaps it
  is not even necessary for the I-Now to change only in a sequential manner
  as in these variants; perhaps it can change along multiple dimensions,
  corresponding to changes across time and changes across space or
  individuals. \citet{Skow09:Relativity}'s relativistic moving-spotlight
  theory, in which individual points in spacetime are ``lit up'' from the
  perspective of points in {\em superspacetime}, seems very much in line
  with such a view.  This also raises important qestions about how these
  dimensions interact: Is temporal change objective or subjective? Is
  subjectivity eternal or temporary?  For related questions on the
  interaction of time and modality, see \citet{Dorr19:Diamonds}.}  Again,
my goal in doing so is merely to illustrate some of the possibilities that
the theory leaves open. The availability of multiple distinct
interpretations should not be surprising given the analogy and
interrelation with the A-theory.  It is also clear that some of these
versions are much more solipsistic than others, or, at least, fit the
negative connotations of solipsism more than others.

Moreover, in earlier work on theories resembling the $\alpha$-theory,
scenarios are often sketched that fit much better with some of these
versions than with others.  Usually, this is done without much discussion
of why the author prefers such a version or even of what the alternatives
might be.  This has the effect of opening up the theory to criticisms that
another version of the $\alpha$-theory might have avoided.  Consider the
following passage by~\citet[page 51]{Hare09:On} (discussing a thought he
had as a child), corresponding to a single-individual-overall theory:

\begin{quote}
  Isn't it amazing and weird that for millions of years, generation after
  generation of sentient creatures came into being and died, came into
  being and died, and all the while there was this absence, and then one
  creature, CJH, unexceptional in all physical and psychological respects,
  came into being, and POW! Suddenly there were present things!
\end{quote}
Later on,~\citet[page 83]{Hare09:On} considers a type of reincarnation:
\begin{quote}
  Is it necessary that only one person ever have present experiences?
  Again, the natural thing is to say no. Egocentric presentism gives me
  conceptual resources to imagine being one sentient creature, and then,
  later, being another sentient creature.  So (recall Nagel's ``fantasy of
  reincarnation without memory'') I can imagine that, after a lifetime of
  oblivious egg consumption, I die a happy philosopher, then find myself in
  a cage eighteen inches tall by twelve inches wide, my beak clipped to its
  base.  This need not involve imagining that CJH dies a happy philosopher
  and then becomes a battery chicken. It may only involve imagining that
  after CJH's death there are again present experiences, and they are the
  experiences of a battery chicken.  Once again this is a real, real nasty,
  metaphysical possibility.

  So ``the one with present experiences'' is a definite description that
  may be satisfied by different things at different times. Like all such
  descriptions, it behaves as a {\em temporally nonrigid referring term}.
\end{quote}
Similarly,~\citet[page 366]{Valberg13:Temporal} writes:
\begin{quote}
We can, however, give sense to the possibility that a human being other
than JV in the past was ``me,'' or that a human being other JV [sic] might
be ``me'' in the future. That is, it makes sense experientially (as a way
things might be or develop from within my experience) that, in the past, a
human being other than JV occupied the position at the center of my
horizon, or that a human being other than JV will occupy this position in
the future.
\end{quote}

Again, the main point here is to make clear how many possibilities the
$\alpha$-theory leaves open and thereby to prevent overly specific
interpretations.  The discussions in the remainder of the paper generally
apply to all of the above versions of the $\alpha$A-theory.  A reader who
wants to keep just a single version in mind might focus on, for example,
personalized presentism or a personalized moving spotlight theory, with a
single individual overall.

\section{Revisiting arguments in favor of the A-theory}
\label{se:favor}

In this section, I will revisit some well-known arguments in favor of the
A-theory.  \ref{su:simpliciter} concerns the argument from presence {\em
  simpliciter} and~\ref{su:appropriate} concerns the argument from the
appropriateness of sentiments such as those expressed by ``Thank goodness
that's over!''  In both cases, the argument will be shown to support the
$\alpha$A-theory more strongly than the $\beta$A-theory, because the
argument supports a distinguished {\em I} just as it supports a
distinguished {\em Now}.  Whether these arguments are indeed effective
against the B-theory is not the topic of this paper, so I will not review
responses that B-theorists may give to these arguments here.

\subsection{Presence {\em simpliciter}} 
\label{su:simpliciter}

Arguably the most basic argument in favor of the A-theory is that of ``the
presence of experience.''  Many have made such an argument; a good
exposition of one is given by~\citet{Balashov05:Times}.  The argument is
that my current experience of writing this paper is {\em present} (or {\em
  occurs}\footnote{\citet{Balashov05:Times} uses ``presence'' and
  ``occurrence'' to refer to different concepts, but it seems to me that
  others have used ``presence'' to refer to a concept that is closer to
  Balashov's ``occurrence''.  In any case, this latter concept is what I am
  after, and I hope that the use of ``{\em simpliciter}'' makes this
  clear.}) in a way that my going through security at the airport yesterday
is not present.  This is not to be taken as a relative statement; everyone
will agree that the writing experience at 5:50pm on March 18, 2019 is
present {\bf at 5:50pm on March 18, 2019} in a way that the airport
security experience at 8:15am on March 17, 2019 is not present {\bf at
  5:50pm on March 18, 2019}.  Rather, the writing experience seems present
in an {\em absolute} sense that does not require the boldface phrases, and
this is referred to as presence {\em simpliciter}.

I argue that, if we are to entertain such a notion, for it to be at all
palatable, it must be personalized, for the following reason.  Just as my
earlier airport security experience is not present {\em simpliciter},
neither is David's experience of eating breakfast in Australia present {\em
  simpliciter}, even if this event happens to take place at the same
time.\footnote{There is, of course, the question of what ``at the same
  time'' even really means given that in special relativity, simultaneity
  depends on the frame of reference.  I will discuss relativity later; for
  the purpose of the current argument, we may assume a Newtonian universe.}
Let me first attempt to explain what I mean by this, and then argue for it.
In order to clarify what I mean, it is tempting to write that David's
breakfast experience is not present {\em simpliciter} {\bf to me}.  But to
do so would undermine the argument, in the exact same way that it would
undermine the purely temporal version of the argument to say that my
airport security experience is not present {\em simpliciter} {\bf right
  now}.  In the latter sentence, ``{\em simpliciter}'' is clearly at odds
with the indexical ``{\bf right now}.''  The exact same is true about the
juxtaposition of ``{\em simpliciter}'' and ``{\bf to me}.''  If an
experience takes place {\em simpliciter}, then to capture this we should
not add any relativizing indexical phrases.

Moreover, it seems that only an experience can be present {\em simpliciter}
in this way.\footnote{\citet[pages 326-327]{Merlo16:Subjectivism} makes a
  similar point.}  For example, it is not at all clear to me what it would
mean for a chair to {\em itself} be present {\em simpliciter}.  My {\em
  experience} of a chair -- visual, tactile, and the result of significant
cognitive processing -- can be present {\em simpliciter}.  Such an
experience is the kind of thing that can have the ``liveness'' that past
and future experiences do not, and that others' experiences do not.  But I
cannot imagine what it would mean for the chair to {\em itself} be ``live''
in this way.  If we are willing to be a bit loose with our language, in
most cases it will not cause confusion to, as a shorthand, say that the
chair is present {\em simpliciter} when we really mean to refer to my
experience of the chair.  But if we are being strict, the experience is not
the chair itself.  Moreover, it seems that an experience can only be had by
a single person\footnote{I use ``person'' here, and throughout, in a broad
  sense; presumably animals and perhaps artificial intelligence can
  similarly have experiences.  Also, in common parlance, of course two
  people can ``share an experience,'' but I use ``experience'' here more
  narrowly in its phenomenological sense.} at a single time,\footnote{Along
  the same lines,~\citet[page 49]{Hare09:On} describes the distinguished
  nature of his current experience and emphasizes that it is an
  easy-to-make ``big mistake'' to extend this to other current
  experiences. \citet{Hare10:Realism} presents an argument with strong
  similarities to the one presented here.  Finally, at the end of his
  paper,~\citet{Skow09:Relativity} also discusses the vivid nature of
  present experiences and argues that a local spotlight shining on a single
  individual explains this just as well as a global one (though he does not
  argue that it actually explains it {\em better}).}  and it does not seem
that two distinct experiences, corresponding to different individuals
and/or times, can be co-present {\em simpliciter} in this way.  So, if
anything, the argument would suggest the existence of a metaphysically
distinguished (I, Now) pair.

Is this argument equivocating between ``presence'' in the temporal sense
and ``presence'' in the experiential sense?  Indeed both meanings of the
word seem to play a role, and I believe that this is revealing rather than
misleading.  Insofar as the current moment in time has a ``liveness'' that
other moments do not, it has it only through my own experience; the same
moment elsewhere, even if experienced by someone else, lacks this liveness
just as a past moment here, even if experienced by me, lacks it.  In this
way, the two meanings of the word are inextricably linked.  \citet[page
100]{Hare09:On} similarly argues that it is in fact advantageous that the
word ``present'' has multiple readings.

It is also important here not to be misled by how we use language.  The
sentence ``David is eating breakfast'' is, in a sense, simpler than ``I
went through airport security yesterday morning.''  Both sentences
explictly refer to their subject (``David'' and ``I''), but only the latter
needs to explicitly refer to when the event took place (``yesterday
morning'') in order to place it in time.  So the first sentence has a type
of simplicity that the second one lacks; we could add ``now'' to the
former, but it is not needed.  On the other hand, dropping ``I'' from the
second sentence leaves it grammatically mangled.  From this asymmetry
between ``I'' and ``now'' one might be tempted to conclude that the word
``simpliciter'' more naturally corresponds to what is happening {\em now}
-- since the word ``now'' is usually not needed for sentences concerning
the present -- than it would correspond to what is happening to {\em me} --
since a word such as ``I'' or ``me'' is usually needed for a sentence
concerning the first person.

However, I would argue that the significance of this asymmetry is not
metaphysical, but rather entirely linguistic.  So many of our spoken
sentences concern the present that, pragmatically, it would be inefficient
to require adding a word like ``now'' to all these sentences.  On the other
hand, usually a conversation concerns multiple actors, so it is important
to make it clear who is the subject in each sentence.  To make this clear,
consider a different context: my planner.  In my planner, I write entries
such as ``attend faculty meeting at noon.''  It would be an inefficient use
of my time to add ``I'' (or ``I will'') to the beginning of the sentence,
because I would have to do so for almost every entry in my planner!  In
contrast, naturally, each of my planner entries {\em must} have a time
associated with it; after all, if the event were happening right now, I
would not have to add an entry to my planner.  So, in the context of my
planner, the roles that subject and time play in the pragmatic issue at
hand are reversed: the former is generally implicit but the latter is
not.\footnote{For additional discussion of the linguistic asymmetry between
  time and space, and how this asymmetry is driven by pragmatic concerns in
  communication, see~\citet{Butterfield84:Seeing}.}  This appears to
confirm that the asymmetry is due to pragmatic reasons.

\subsection{The appropriateness of wanting things to (not) be past} 
\label{su:appropriate}

Another well-known argument~\citep{Prior59:Thank,Zimmerman07:Privileged} in
favor of the A-theory (and presentism in particular) concerns the
appropriateness of statements such as ``Thank goodness that's over!''
Here, ``that'' might refer to something like a headache the speaker was
experiencing.  It is often argued that the B-theory does not provide the
resources to capture the full significance of this statement.  Prior argues
that the meaning of such a statement is not that it is good that the
headache takes place at a point in spacetime earlier than the point at
which the statement is uttered; in his words, ``Why should anyone thank
goodness for that?''  Instead, what the statement is getting at is that the
headache is simply {\em over}, and the A-theory provides the resources to
capture this.  But one might similarly argue in favor of the
$\alpha$-theory, for example appealing to the appropriateness of statements
such as ``Thank goodness that is not happening {\em to me}!'' This is
closely related to the question of whether self-bias could be
metaphysically justified, as studied by~\citet{Hare07:Self,Hare09:On}.  The
$\beta$A-theorist is likely to complain that the analogy is not apt,
because the second statement merely reflects a selfish disposition rather
than something more fundamental.  It is not clear to me why the same could
not be said of the first statement, that the statement merely reflects the
speaker's callousness towards her past self.  To avoid this criticism,
perhaps one can make the first statement about someone else (``Thank
goodness John's headache is over!''), but, and I believe this is telling,
the argument seems to lose force with this move.

Let us explore this in a bit more depth. Suppose all headaches last exactly
one or two days with no ill effects afterwards, and consider the following
two statements:
\begin{itemize}
\item[$S_1$:] Thank goodness John's headache, which started yesterday,
  ended yesterday as well, rather than continuing into today.
\item[$S_2$:] Thank goodness John's headache, which started the day before
  yesterday, ended the day before yesterday as well, rather than continuing
  into yesterday.
\end{itemize}
Here, we imagine caring a great deal about John and preferring him not to
suffer.  Under the $\beta$A-theory, one would expect $S_1$ to have a
significance not shared by $S_2$, as the former concerns a difference in
what is happening {\em now}, whereas the latter concerns a difference that
is in any case entirely in the past.  It is not clear to me that such a
difference in significance is really there. Is it not just as reasonable to
appreciate that John did not suffer yesterday, as it is to appreciate that
he is not suffering today?

Yet, one may have an intuition that indeed, $S_1$ has a significance that
$S_2$ does not.  I believe that the likely grounds for this intuition are
not germane to the issues under discussion here, and we can modify the
scenario to remove these grounds.  First, in the first situation, if John
were still having a headache, I might feel compelled to try to {\em do}
something to alleviate his suffering.  However, this is easily addressed by
postulating that it is common knowledge that I can do nothing of the sort.
Second, if John is in my immediate environment and I see him suffering,
this may cause me to suffer as well, for example due to the mirror neurons
in my brain.  But this is merely returning us to an example where I myself
suffer, which is precisely what we were trying to avoid by introducing
John.  Hence, we should postulate that John is somewhere else entirely.

To make all this concrete, suppose that John has decided to go on a
two-month retreat in a faraway country.  He will not communicate until he
gets back.  Halfway into his retreat, I realize that around this time of
year, he always gets a headache, which may last one or two days.  I care
for him and so I hope that it is just a one-day headache this time.  But I
will not find out until he comes back and tells me.  Imagining this
scenario, I do not find myself concerned specifically about whether his
headache happens to be taking place right now, or not.\footnote{In this
  example, there is nothing to synchronize John's experience with mine; his
  life is unfolding in parallel to mine and it is hard to see why it would
  matter which events are contemporaneous.  As we will discuss
  in~\ref{su:relativity}, we can make the example even more extreme by
  having John fly far off into space somewhere, so that, as far as the
  theory of relativity is concerned, there really is no absolute answer to
  the question whether his headache is taking place at the same time as my
  current experience.  If so, caring about simultaneity seems to require a
  very strong commitment to the $\beta$A-theory, as it requires that there
  be an additional fact about simultaneity over and above the theory of
  relativity that is important for what we should care about, even though
  no physical measurement could ever tell us whether two events actually
  were or were not simultaneous in this sense.}

Hence, given that the scenario is set up appropriately, I remain
unconvinced that there is any significant difference between $S_1$ and
$S_2$, and this seems to deal a blow to the $\beta$A-theory.  Naturally,
the $\beta$B-theory avoids this blow; but I believe the $\alpha$A-theory
also avoids it, in that John today is just as much ``outside the I-Now'' as
John yesterday, because I am not John.  In fact, compared to the
$\beta$B-theory, the $\alpha$A-theory does a better job explaining why
something about the example seems to change when I myself am brought into
it.  That is, if we replace ``John's'' with ``my'' in the statements above
to obtain $S_1'$ and $S_2'$, then it does seem that $S_1'$ has a
significance that $S_2'$ does not.  $S_2'$ is not an unreasonable statement
-- it makes sense to appreciate having suffered less than one might have,
just as it makes sense to appreciate someone else suffering less than he
might have -- but only $S_1'$ concerns the immediate presence or absence of
suffering, which is the vivid characteristic that imbues ``Thank goodness
that's over!'' examples with their intended significance.\footnote{Some of
  this is reminiscent of \citet{Turri13:That}'s ``That's outrageous!''
  example.  Turri argues that just as the appropriateness of statements
  such as ``Thank goodness that's over'' can be used to support presentism,
  the appropriateness of statements such as ``That's outrageous!'' can be
  used to attack it, because it seems perfectly legitimate to be outraged
  by, say, a past genocide.  I consider it telling that ``Thank goodness
  that's over!'' examples typically involve oneself and ``That's
  outrageous!'' examples typically involve others; this may well be what is
  driving the difference in conclusions from these examples.}

Indeed, both~\citet{Suhler12:Thank} and~\citet{Greene15:Against} report on
an experimental study by~\citet{Caruso08:Wrinkle} in which subjects were
asked what would be fair compensation for a particular task.  The study
found that when subjects were asked to imagine themselves doing the task in
the future, they felt that they should be compensated significantly more
than when they imagined themselves doing the task in the past; but this
effect disappeared when they were asked to imagine someone {\em else} doing
it.  \citet{Suhler12:Thank} take this to invalidate the ``Thank goodness
that's over'' argument, and~\citet{Greene15:Against} argue for complete
temporal neutrality in making decisions.  (The argument for temporal
neutrality is worked out in detail in~\citet{Sullivan18:Time}.  \citet[page
61]{Hurka96:Perfectionism} argues that temporal neutrality is appropriate
for certain non-hedonic goods, but is convinced that it is not for avoiding
pain, by the example from~\citet[page 165]{Parfit84:Reasons} that we would
prefer a more painful operation in the past to a less painful one in the
future.)  The analysis above suggests that while indeed, the results of
the~\citeauthor{Caruso08:Wrinkle} study cast doubt on whether the ``Thank
goodness that's over'' argument effectively supports the $\beta$A-theory,
they are perfectly consistent with this argument supporting the
$\alpha$A-theory.

\section{Revisiting arguments against the A-theory}
\label{se:against}

In this section, I will revisit some well-known arguments against the
A-theory. \ref{su:relativity} concerns the argument from special
relativity,~\ref{su:direction} concerns the argument that the direction of
time may be a local matter,~\ref{su:rate} concerns the argument that asks
for the rate at which time passes, and~\ref{su:travel} concerns the
argument from time travel and G\"odelian universes.  In all cases, the
$\alpha$A-theory will be shown to avoid most of the bite that these
arguments inflict on the $\beta$A-theory, roughly because the arguments
hinge on the fact that the Now is global in nature -- that is, it stretches
across all of space.  Because the I-Now is local in nature, the arguments
are ineffective against the $\alpha$A-theory.

\subsection{Special relativity}
\label{su:relativity}

Einstein's theory of relativity has often been invoked to criticize the
A-theory.  Unlike in a Newtonian universe, in the special theory of
relativity, simultaneity is not absolute; rather, whether two events are
simultaneous depends on the reference frame.  But if there is no absolute
simultaneity, then how can there be an absolute Now?  Special relativity
can also be used to cast doubt on specific arguments in favor of the
A-theory -- or at least, the $\beta$A-theory.  For example, let us modify
the example from~\ref{su:appropriate} by putting John on a faraway planet,
so that whether his headache is earlier or later than our own time depends
on the reference frame.  This seems to make it difficult to hold the
position that, in order to know how we should feel about John's headache,
it is important to know whether it is in the past or in the future.  Now,
perhaps there may still be a separate, absolute sense in which John's
headache is in the past, even if this is not implied by the theory of
relativity.  But if there is not, this poses a problem for using the
``Thank goodness that's over!'' argument in support of the $\beta$A-theory
-- but, importantly, not for using it in support of the $\alpha$A-theory,
because, as discussed in~\ref{su:appropriate}, in that case the argument is
only made about one's own pains rather than those of someone on a faraway
planet.  Still, we must investigate the implications of relativity for the
$\alpha$A-theory more broadly.

Some (e.g.,~\citet{Markosian04:Defense}) have argued that, in fact, a
philosophically austere version of the theory of relativity could explain
the empirical evidence without implying that there is no absolute
simultaneity.  The relation of absolute simultaneity could be added on top
of the theory of relativity.  For example, one might suppose that there
exists a distinguished frame of reference that determines which events are
absolutely simultaneous.  Positing such a distinguished frame seems a
rather awkward and inelegant addition to the theory, one that is rather
contrary to the spirit of the theory of relativity and perhaps more in line
with older theories of a stationary aether.
But,~\citet{Zimmerman07:Privileged} has argued that such an addition to the
physical theory is no different in kind from the addition of a
distinguished Now in the first place.  That may be so, but it is a further
addition, and it seems that, for the sake of parsimony, each addition
should at least count against the resulting theory.  The analogy is also
imperfect.  It can at least be argued that we know when the Now is; in
contrast, it is not clear whether and how we could ever know what the
distinguished frame of reference is.  \citet{Zimmerman11:Presentism}
discusses and responds to all these concerns in far more detail than I can
do here, and argues well that they are not fatal to the $\beta$A-theory,
but it is clear that at least they pose significant challenges.

In any case, the above arguments only concern the $\beta$A-theory.  In the
$\alpha$A-theory, there is no need for any observer-independent
simultaneity at all.  While the Now in the $\beta$A-theory must be global
-- in the sense that everywhere in the universe, there are events happening
Now, thereby introducing an observer-independent simultaneity relation
across all of space -- the I-Now in the $\alpha$A-theory is local.  The
precise nature of this locality -- for example, whether the I-Now is
spatially extended -- does not matter much for the arguments at hand; what
matters is that the I-Now is associated with an observer, and that that
observer can be localized in spacetime.  Specifically, this ties the I-Now
to the frame of reference associated with that observer;\footnote{The
  definition of what constitutes a frame of reference varies.  Here, we
  consider a frame of reference to be determined purely by its state of
  motion, rather than to also include a coordinate system.} if so desired,
simultaneity could be determined based on this frame of reference according
to the theory of relativity.  For that matter, no notion of simultaneity
across space is even required for the theory to make sense.  While the
$\beta$A-theory necessitates such a notion -- whatever is happening Now
across space must be simultaneous, in an objective sense -- it does not
seem to pose any problem for the $\alpha$A-theorist to hold that there is
no absolute notion of simultaneity.  As far as the $\alpha$A-theorist is
concerned, we can define a notion of simultaneity for convenience, for
example the one based on the theory of relativity and the distinguished
frame of reference corresponding to the I-Now as just suggested, but none
is truly needed.  In fact, the problems that the theory of relativity poses
for the A-theory have already led to at least one proposal similar to the
$\alpha$A-theory, namely \citet{Skow09:Relativity}'s relativistic spotlight
theory,\footnote{In earlier work, \citet[page 18]{Stein68:On} hints at a
  similar theory when he contemplates what would result from an argument by
  \citet{Putnam67:Time} if one tried to preserve a different intuition
  about the relationship between what is present and what is real.  It is
  not clear whether he intends at all to defend such a theory.}  in which
the spotlight shines locally, not globally.\footnote{\citet{Hare10:Realism}
  and~\citet[page 48]{Hare09:On} also make some of the points that I made
  in this subsection.  \citet{Fine05:Tense,Fine06:Reality} similarly gives
  a detailed discussion of what, for the realist, should replace the role
  of times when we take into account special relativity, and concludes that
  most plausibly frame-time pairs should take their role, in combination
  with a nonstandard type of realism in which either realities are indexed
  to different frame-times or reality is fragmented.}

\subsection{The direction of time}
\label{su:direction}

For any version of the $\beta$A-theory in which time flows, there needs to
be an objective {\em direction} in which time flows.  Presumably, it flows
from what we perceive as the past to what we perceive as the future.  But
if the laws of physics are invariant to time reversal, then these laws do
not naturally provide such a direction.  It is commonly held that what we
perceive as the direction of time is tied to the entropy gradient, and that
this entropy gradient may well be reversed in other parts of spacetime.  If
so, we may imagine a Doppelg\"anger being that is otherwise very much like
ourselves, living its life in such a part, backwards in time from our
perspective~\citep{Williams51:Myth,Maudlin02:Remarks}.  The Doppelg\"anger
would presumably think that {\em we} have it backwards, that the direction
of time's flow is opposite from what we think it is.  So what gives us
reason to believe that we are the ones to have it right?  A key issue here
is that presumably, the $\beta$A-theory requires time to flow in the same
direction everywhere; the direction should be {\em globally}
consistent.\footnote{The Now is not localized under the $\beta$A-theory, so
  that there is a single Now across space; but if it moves in one direction
  in one location and in the opposite direction elsewhere, it is hard to
  imagine that after moving in these opposite directions it remains the
  {\em same} Now across these locations.} It has been argued that we have
no reason to believe that the Doppelg\"anger even has mental states at all,
by virtue of the fact that the way its life proceeds is so unlike the way
ours proceeds~\citep{Maudlin02:Remarks}.  But this seems a rather odd
conclusion, since we have supposed that, {\em mutatis mutandis} for the
difference in direction, the Doppelg\"anger's life is entirely like ours.
For a more detailed discussion of this point and these issues more
generally, see~\citet{Price11:Flow} and references cited therein.

In contrast, the putative existence of persons living in parts of spacetime
with a reversed entropy gradient, living their life backwards in time (from
our perspective), poses no problem for the $\alpha$A-theory. This is
because the I-Now is inherently {\em local} (in both a spatial and a
temporal sense), so it does not matter if the entropy gradient is reversed
elsewhere; all that matters is what the entropy gradient is {\em here} (and
{\em now}), because that is what determines the direction in which the
I-Now moves.  If the I-Now actually tracks a Doppelg\"anger at some point,
it does not appear to pose any problem for the theory for it to then move
in the opposite direction.  (This may pose problems for some of the
specific illustrative versions presented earlier in
Section~\ref{se:versions}, but it poses no problem for the other versions.)
We can view {\em external} time as nothing more than a dimension through
which the I-Now travels.

Taking this to an extreme, we may even imagine a machine that transports
you to another region of space where the entropy gradient is reversed
relative to ours, and that transforms you into a Doppelg\"anger there.  You
will, in some sense, continue your life there uninterrupted, except moving
in the opposite temporal direction.  Of course being transported to another
region of space is likely to be a bit shocking; but, if such scenarios are
possible at all, there seems to be no reason to believe that your
experiences will be any different than they would have been if you had been
transported instead to a region of space that happens to have the same
entropy gradient (and not been transformed into a Doppelg\"anger).
Accommodating this intuition is easy under the $\alpha$A-theory; for
example, the I-Now could simply jump along with you and then start moving
backwards (from our initial perspective). On the other hand, this example
appears problematic for versions of the $\beta$A-theory that require a
globally defined direction of time, because such a theory would have to
conclude that one of the two halves of your life is lived, in an {\em
  absolute} sense, backward.  If we believe~\citet{Maudlin02:Remarks}'s
argument, we would then conclude that you had real mental states in only
half of your life.  This seems to be an odd conclusion.  If near the end of
your life you were transported back to the original spacetime region, the
suggestion that you had not had any real mental states since the original
transportation event would seem utterly bizarre to you!

\subsection{The rate of time's passage}
\label{su:rate}

Opponents of the A-theory (or $\beta$A-theory) have also criticized it as
follows: if the Now moves, what is the rate at which it moves?  It has been
argued that if one says that it moves at $1$ second per second, this poses
a problem for the theory, because one can cancel the units of seconds and
conclude that the rate is simply $1$, and (supposedly) $1$ is not a rate
(e.g.,~\citet{Olson09:Rate}).  Now, the idea that a unitless rate is not a
rate is simply nonsense.  This has been convincingly argued
elsewhere:~\citet{Skow11:On} uses the example of sociologists tracking what
the ``most common birth year'' in the population is.  One would expect the
most common birth year to generally increase by roughly $1$ year every
year, though the rate may be higher or lower than $1$ depending on
demographic phenomena. In any case, the rate is unitless (one might just as
well say the rate is approximately $1$ decade per decade).  The example is
convincing to me, and clearly many other examples of sensible unitless
rates can be provided.  One such example is particularly relevant here: due
to relativity, satellites and astronauts on the International Space Station
age at a slightly different rate than objects and people on the surface of
the Earth.  The amount of time that such a satellite or astronaut
experiences per unit of Earth surface time is a unitless rate.\footnote{One
  might counter that these conditions in fact correspond to different
  units, namely Earth surface seconds and ISS seconds, so that we in fact
  do not obtain a unitless rate.  But this misses the point that a second
  denotes the same amount of aging for the people in each condition.  The
  unitless rate indicates how much faster people in one condition age than
  those in the other, and for this comparison no units are needed.
  Similarly, we need no units to say that one person is 1.2 times as tall
  as another.  That the rate being unitless is meaningful is further
  illustrated by the fact that it can be both above and below 1, because of
  the opposing effects of relative velocity time dilation and gravitational
  time dilation; there is an orbit, about half the radius of the Earth
  above the surface, at which the rate is 1~\citep{Ashby02:Relativity}.
  The rate being 1 at this orbit is not just a meaningless consequence of
  how we defined the units; it is the orbit at which astronauts age equally
  fast as those on the surface.} This example actually seems to pose a more
serious problem for the answer that time moves at ``$1$ second per second''
-- if the idea is to think of time as moving globally rather than just
locally, then in just {\em whose} seconds are we measuring this rate?  In
any case, a weaker version of the original criticism seems to hold up: the
question only allows uninformative answers.  The answer that it moves at
``$1$ second per second'' seems tautological.  We could instead introduce
the concept of {\em supertime} to track the Now's motion through time, so
that at different points in supertime, the Now is at a different time. (For
a detailed discussion of the metaphor of supertime,
see~\citet{Skow12:Why}.)  Then, we can ask how many seconds pass per
supersecond.  However, there seems to be every reason to simply define the
supersecond so that the answer becomes ``$1$ second per supersecond,''
which remains uninformative.

In the $\alpha$A-theory -- or, at least, in versions of it where the I-Now
moves along with a person through time (see Section~\ref{se:versions}) --
the question of how fast the I-Now moves does not pose such
problems. First, the fact that on a space station, a different amount of
time is experienced to pass no longer poses any problem, because the I-Now
is local, so there is no requirement that time passes at the same rate
everywhere.  Moreover, the question of how fast the I-Now moves can have
more interesting answers.  In the relativistic example above, it is natural
to respond that the I-Now moves at a different rate when it is associated
with an astronaut in orbit than it does when it is associated with a person
on the surface.  Alternatively, let us put relativity aside for a moment
and focus on the I-Now's experiential aspect instead. One might reasonably
hold that the I-Now moves through external (i.e., clock) time at a
different rate when it is associated with a person who is under anesthesia
than it does when it is associated with someone who is highly alert.
  
If we allow ourselves to speculate, a computational\footnote{It is
  important to hold a sufficiently broad view of ``computation'' here; such
  broad views are common among those working on the theory of computation.
  Alternatively, and less ambitiously, the reader may just view this as a
  suggestive analogy to the clock speed of a computer.}  theory might be
used to unify these two examples: consider a person's ``clock speed'' --
the number of mental operations, according to some suitable definition, per
(Earth surface) second -- and take this to determine the rate at which the
I-Now moves.  Specifically, let us define a supersecond so that there is
always exactly one mental operation per supersecond.  Then, the number of
(Earth surface) seconds per supersecond -- which is just the reciprocal of
the clock speed defined above -- will vary in the different scenarios
above, in a way that conforms with our intuitions.  Focusing on Earth
surface seconds per supersecond (regardless of the location of the person)
simultaneously addresses both the relativistic and the experiential
components of the scenarios, and also allows us to handle mixed cases, such
as a space station inhabitant who is under anesthesia. In such a case, the
number of mental operations per Earth surface second can be written as the
number of mental operations per space station second, multiplied by the
number of space station seconds per Earth surface second, thereby
separating out the experiential and relativistic components, respectively.
This shows that these two components are compatible.  Per the theory of
relativity, there is nothing special about Earth surface seconds, as
opposed to space station seconds or Mars surface seconds; they are just
different ways to measure external time.

Supertime, so defined, perhaps more naturally corresponds to our sense of
passage, leaving regular time (as tracked by clocks) in the more modest
role of a dimension through which we happen to pass, as noted earlier.
That is, this notion of supertime would allow us to give metaphysical
meaning to the idea of time passing more or less quickly from a subjective
viewpoint.  Of course, this view may conflict with other intuitions that we
have developed.  In our ordinary experience of time, relativistic issues do
not come into play, and our waking experience of how fast time passes is
usually fairly stable.  Given this, we tend to conceive of time as
objective, and treat any variance in how we perceive its passage as a mere
error in estimation.  For the current purpose, I believe such intuitions
are misleading.  The following two examples are intended to illustrate that
it is in fact quite natural to assign primary importance to the notion of
supertime as defined here.  In each of them, we will imagine a choice
between two alternatives that result in you having different amounts of
time but equal amounts of supertime left in your life.  I argue that you
should be (close to) indifferent between the options in both scenarios.

{\em Example 1.}  It is the year 2400, and you are part of a group of
people on a lifelong space voyage.  The group is about to split up into two
subgroups that will take separate spacecraft.  It is common knowledge that
the two subgroups will never communicate again, either with each other or
with the people left on Earth.  You get to choose in which subgroup you
will be.  They are indistinguishable, except the two spacecraft will move
to orbits around different massive bodies, with different relativistic time
dilations.  If you choose to be on spacecraft 1, your life will therefore
be shorter in Earth time than it would be on spacecraft 2.  As a result,
your first reaction may be that you would prefer to be on spacecraft 2.
But, I argue, upon closer inspection there is little reason for this.  This
is because, to make up for the shorter amount of Earth time in your life on
spacecraft 1, correspondingly more events will happen per unit of Earth
time on spacecraft 1.  You would experience entirely similar lives on the
two spacecraft, with equally many interesting events taking place on both.
If it were possible to communicate from Earth to the spacecraft, you might
prefer being on spacecraft 2 because (for example) more papers, books, and
movies would be produced on Earth and sent to spacecraft 2 for your
consumption during your life.  But we have assumed that such communication
is impossible.  As far as I can see, there does not seem to be any
compelling reason to have a preference about on which spacecraft you
continue your voyage.

{\em Example 2.} It is again the year 2400, but this time we will stay on
the surface of the Earth.  After a long and happy life, you have
regrettably contracted an incurable disease that, if left untreated, will
kill you almost immediately.  Unfortunately, the only possible treatments
will put you in a type of comatose state until your death.  You will,
however, have wonderful dreams in this state.  Due to secrecy issues, your
friends and family will never be made aware of your predicament.  There is
no chance at all that any new treatment will become available during the
remainder of your life.  You have a choice between medications $M_1$ and
$M_2$.  Compared to $M_1$, $M_2$ would keep you alive for twice as long,
but would allow your brain to process at only half the rate.  Your first
reaction may be that you would prefer to receive $M_2$.  But again, I
argue, upon closer inspection there is little reason for this.  Because of
the difference in brain processing rates, you would have equally many
wonderful dreams under the two medications.  If your friends and family
could visit you in your comatose state, you might prefer for them to have
that option for a longer or shorter period of time, but we have ruled this
out.  If you had hopes that scientists could develop a cure, you would
prefer $M_2$ to give the scientists more time, but we have also ruled this
out.  As far as I can see, there does not seem to be any compelling reason
to have a preference about which medication you receive.

In summary, to the extent that the question about the rate at which the Now
moves poses a problem for the $\beta$A-theory, it does not pose this
problem for the $\alpha$A-theory, since for the latter the answer to the
question need not be tautological.

\subsection{Time travel and G\"odelian universes}
\label{su:travel}

A final criticism of the ($\beta$)A-theory is that it does not make much
sense of time travel scenarios. Following~\citet{Lewis76:Paradoxes}, it
seems natural to distinguish between {\em external} time and the time
traveler's {\em personal} time.  But if one takes external time seriously
in the metaphysical sense, as would be expected of a $\beta$A-theorist, it
would appear one cannot simultaneously do the same for personal time.
This, in turn, necessitates unintuitive attitudes towards time travel.  The
following passage by~\citet[page 333]{Sider05:Traveling} illustrates this
perfectly.

\begin{quote}
  But if personal time bears little similarity to external time then
  ``personal time'' is merely an invented quantity, and is misleadingly
  named at that.  That I will view a dinosaur in my personal future amounts
  merely to the fact that I once viewed a dinosaur, and moreover that this
  is caused by my entry into a time machine. Since this fact bears little
  resemblance to the facts that constitute a normal person's genuine
  future, I could not enter the time machine with anticipation and
  excitement at the thought of seeing a dinosaur, for it is not true that I
  am {\em about} to see a dinosaur, nor is the truth much {\em like} being
  about to see a dinosaur. If anything, I should feel fear at the thought
  of being annihilated by a device misleadingly called a ``time
  machine''. The device causes it to be the case that I once viewed a
  dinosaur, but does not make it the case in any real sense that I {\em
    will} view dinosaurs.
\end{quote}
Perhaps there is a way out of this conclusion for the $\beta$A-theorist,
but I cannot see it.  Or perhaps she is willing to bite the bullet and
accept the conclusion that (at least backward) time travel is to be avoided
at all cost.  In any case, the $\alpha$A-theorist avoids this issue.  For
her, personal time is what is taken seriously, and she can legitimately
look forward to -- if this is in fact something to look forward to -- her
encounter with a dinosaur.\footnote{Well, she may still hesitate, to the
  extent that it is not obvious that the presence of experience, the I-Now,
  will follow her through the time machine rather than go somewhere else.
  As an example that illustrates this ambiguity, it may be one of these
  unmarketable time machines that also leave behind a badly burned body,
  apparently alive for a few more seconds, where the traveler entered the
  time machine. (See \citet[page 58]{Hare09:On} for a similar example.)
  But at least her believing that it will follow her back in time (rather
  than transitioning to a different person at the same time, or staying
  with a burnt body) would not cause any inconsistency with her other
  beliefs.}

Closely related to the issue of time travel is that of G\"odelian universes
that cannot be given a global temporal ordering.  The theoretical
possibility of such universes perhaps poses a problem for some versions of
the $\beta$A-theory.  The $\alpha$A-theory, however, does not require any
global temporal ordering.  For versions of the $\alpha$A-theory with a
moving I-Now, one may yet worry if such universes do not create different
problems.  For example, \citet{Dieks06:Becoming} discusses an example
by~\citet[pages 141-142]{Reichenbach58:Philosophy} in which a person loops
around to meet his earlier self again at a particular point in spacetime.
Dieks, who argues for a B-theoretic notion of local becoming, argues that
this example illustrates that even a local type of spotlight is
problematic.  He argues that when the spotlight shines on the region in
spacetime where the younger and older versions of the person meet, there
must in fact be two distinct spotlights, one that will travel with the
younger version and one that will travel with the older version.  Then, the
spotlight associated with the younger version loops around as that version
becomes the older version, eventually reaching the same region again. By
the same reasoning as before, we will again need two spotlights at this
point. But the other spotlight, the one that was initially associated with
the older version, is not available for the task, being meanwhile
associated with an even older version. So we will need a third spotlight,
and so on ad infinitum, which seems problematic.

But it is easy to find an escape from Dieks' argument.  The fact that the
two versions of the person are (roughly) at the same point in spacetime
does not imply that the spotlight shines on them simultaneously {\em in the
  supertime sense}.  That is, the ``same'' spotlight might earlier (in
supertime) light up the younger version only (i.e., that version's
experience at that point) and later (in supertime) the older version only.
Hence, there is no need to introduce additional spotlights when the meeting
point is reached.  This illustrates one advantage of associating the
spotlight with person-stages (I-Now) rather than with small regions of
spacetime (Here-Now): even though the younger and the older version are
both in (roughly) the same location at the same time, they correspond to
different person-stages.  This requires, of course, that in this type of
scenario we associate the I-Now with a person-stage (where a younger and an
older version of the same person at the same time are still considered
separate person-stages), rather than with a pair of a person and a time,
which in this case might pick out both person-stages.  This interpretation
of the I-Now in any case aligns better with the other arguments presented
in this paper.  For example, it seems hard to imagine the (simultaneous)
presence {\em simpliciter} of the {\em combination} of both person-stages.
Also, the older person-stage may think, looking at the younger
person-stage, ``Thank goodness I am no longer that immature!''  The idea
that the spotlight was previously (in the supertime sense) associated with
the younger person-stage and now with the older person-stage seems to
capture the significance of this statement well.  Finally (and more
speculatively), if we imagine the brain of the older stage to have slowed
down and no longer to be processing at the rate of his younger self,
associating the I-Now with person-stages would allow us to say that the
I-Now moves at a different rate with respect to external time when
associated with each of these two person-stages.

\section{Conclusion}

Upon inspection, key criticisms of the A-theory are only effective as
criticisms of the $\beta$A-theory, and key arguments in favor of the
A-theory are much more convincing as arguments for the $\alpha$A-theory.
To the extent I have succeeded in showing that A-theorists are rationally
compelled to be $\alpha$-theorists as well, surely many will interpret this
as a significant blow to the A-theory because they consider the
$\alpha$-theory implausible.  Nevertheless, some philosophers may well be
willing to adopt some version of the $\alpha$A-theory (Hare being an
obvious example).  As I emphasized earlier, a detailed discussion of the
relative merits of the $\alpha$A-theory and the $\beta$B-theory is outside
the scope of this paper.  Such a discussion is sure to revisit many
familiar arguments in the philosophy of time and modality (and mind), and
is unlikely to reach a swift conclusion.\footnote{This seems all the less
  likely given that the problem connects to other challenging problems,
  such as the Sleeping Beauty problem -- see,
  e.g.,~\citet{Conitzer15:Can}.}  I do hope to have convinced the reader
that the $\alpha$A-theory will fare better in such a comparison than the
$\beta$A-theory.  The former has an internal consistency that allows it to
escape some of the more damaging criticisms to which the latter has fallen
prey.

\section*{Acknowledgments}

I am thankful to anonymous referees who provided especially thorough and
helpful comments, which significantly improved the paper.

     \begingroup
     \parindent 0pt
     \parskip 1ex
     \theendnotes
     \endgroup

\end{document}